\documentclass[aps,prl,showpacs,twocolumn,amsmath,amssymb,superscriptaddress,footinbib]{revtex4}

\usepackage[english]{babel}
\usepackage{latexsym}
\usepackage{graphics,psfrag}
\usepackage{epsfig}
\usepackage{color}

\def\be{\begin{equation}}
\def\ee{\end{equation}}
\def\bea{\begin{eqnarray}}
\def\eea{\end{eqnarray}}
\def\bi{\begin{itemize}}
\def\ei{\end{itemize}}
\def\bin{\begin{enumerate}}
\def\ein{\end{enumerate}}
\def\la{\langle}
\def\ra{\rangle}

\newcommand{\rmd}{\mathrm{d}}

\newcommand{\hl}{\xi} 

\begin{document}
\title{Anderson localization of solitons}

\author{Krzysztof Sacha}
\affiliation{
Instytut Fizyki imienia Mariana Smoluchowskiego and
Mark Kac Complex Systems Research Center, 
Uniwersytet Jagiello\'nski, ulica Reymonta 4, PL-30-059 Krak\'ow, Poland}

\affiliation{Laboratoire Kastler-Brossel, UPMC, ENS, CNRS;  
4 Place Jussieu, F-75005 Paris, France}

\author{Cord~A.~M\"uller}
\affiliation{Laboratoire Kastler-Brossel, UPMC, ENS, CNRS;  
4 Place Jussieu, F-75005 Paris, France}
\affiliation{
Physikalisches Institut, Universit\"at Bayreuth, D-95440 Bayreuth, Germany} 

\author{Dominique Delande}

\affiliation{Laboratoire Kastler-Brossel, UPMC, ENS, CNRS;  
4 Place Jussieu, F-75005 Paris, France}
 
\author{Jakub Zakrzewski} 

\affiliation{
Instytut Fizyki imienia Mariana Smoluchowskiego and
Mark Kac Complex Systems Research Center, 
Uniwersytet Jagiello\'nski, ulica Reymonta 4, PL-30-059 Krak\'ow, Poland}

\affiliation{Laboratoire Kastler-Brossel, UPMC, ENS, CNRS;  
4 Place Jussieu, F-75005 Paris, France}

\date{\today}

\begin{abstract}
At low temperature, a quasi-one-dimensional ensemble of atoms with attractive
interaction 
forms a bright soliton. When exposed to a weak and smooth 
external potential, the shape of the soliton is {hardly} 
modified, but its center-of-mass motion is affected.
We show that  
in a spatially correlated disordered potential,  
the quantum motion of a bright soliton  displays 
Anderson localization. 
The localization length
can be much larger than the soliton size and could be observed experimentally. 
\end{abstract}


\maketitle

At zero temperature,
cold atoms interacting attractively in a one dimensional (1D) system tend to cluster together, 
forming a bright soliton.
Explicit solutions of the many-body problem can be found in some cases, for example
for contact interactions~\cite{McGuire64}.
Using external potentials, it has been experimentally shown how to put solitons
in motion~\cite{brightexp}. 
What happens to a soliton exposed to a disordered potential? If the potential is
strong, it 
will destroy the soliton. If it is sufficiently
weak and smooth not to perturb the soliton shape, one expects 
the soliton to undergo
multiple scattering, diffusive motion and possibly  
Anderson localization~\cite{anderson1958}. 
Indeed, propagation of waves in a disordered potential 
is profoundly affected by
Anderson localization. Multiple scattering on 
random defects yields exponentially localized density profiles and
a suppression of the usual diffusive transport associated 
with incoherent wave scattering~\cite{lee1985}.
In 1D Anderson localization is a ubiquitous phenomenon~\cite{vantiggelen1999},
which has been recently observed for cold atomic matter waves~\cite{Billy2008}.
It is important to understand
how it is modified when interactions between particles
are taken into account. 

We consider a Bose-Einstein condensate  
in a quasi-1D geometry. 
Within mean-field theory, it is described by the Gross-Pitaevskii energy functional  
\be
\label{GPE}
E=\int dz \left[\frac12 |\partial_z \phi|^2+\frac{g}{2} |\phi|^4 -\mu |\phi|^2\right]  
\ee
in units of $E_0=4m\omega_\perp^2a^2$,
$l_0=\hbar/2|a|m\omega_\perp$, and $t_0=\hbar/4a^2m\omega_\perp^2$ for energy,
length and time, respectively. Here, $\omega_\perp$ denotes the transverse harmonic 
confinement frequency, $a$ the atomic $s$-wave scattering length, and $\mu$ the chemical potential. 
The cases of repulsive and attractive atomic interaction are covered by $g=\pm 1$.  

{The dynamics is extremely different in both cases, so that we}
first discuss the case of attractive interaction, $g=-1$. The ground state of \eqref{GPE} is 
the bright soliton  \cite{zakharov} 
\be
\phi_0(z-q)=\sqrt{\frac{N}{2\hl}}\frac{e^{-i\theta}}{\cosh[(z-q)/\hl]}, 
\label{bs}
\ee
normalized to the total number of particles $N$. The chemical potential
is  $\mu=-N^2/8$ and the soliton width is $\hl=2/N.$  
This ground-state solution has an arbitrary center-of-mass (CM) position $q$ 
and an arbitrary global phase $\theta$ that spontaneously break the 
translational and the 
$U(1)$ gauge symmetry of the energy functional \eqref{GPE}, respectively. 
These degrees of freedom appear as zero-energy modes of Bogoliubov theory, 
and their quantum dynamics requires special attention~\cite{lewenstein,dziarmaga04}. 
 
The energy functional \eqref{GPE} is no longer translation invariant when 
a potential term $\int dz V(z) |\phi|^2$ is added. If {$V(z)$}  
is sufficiently weak and smooth, the soliton shape remains unchanged {to lowest order} {in $V$}, 
and only its {CM} position $q$ is affected. In this Letter, we show that 
the quantum dynamics of  
$q$ {in a disorder potential} leads to Anderson localization of the soliton,
over a localization length possibly longer than the soliton size.

An intuitive picture emerges using a simple ansatz in terms of collective coordinates 
(cf.\ \cite{Gaul2009} in the context of Bloch
oscillations). Inserting 
$\phi(z;q,P_q) = e^{iP_qz/N}\phi_0(z-q)$ into  \eqref{GPE}
leads to an effective quantum Hamiltonian 
\be
\hat{H}_q = \frac{\hat{P}_q^2}{2N}+\int dz\ 
V(z)\ |\phi_0(z-\hat{q})|^2.
\label{bsenergy}
\ee
It describes the entire soliton as an object of mass $N$ evolving in
an effective potential 
$\int dz V(z)|\phi_0(\hat{q}-z)|^2$ 
that is the convolution of the bare potential with the soliton density. 

This simple ansatz yields no information on the remaining degrees of freedom. 
Therefore, we apply a more complete analysis, 
expanding the energy functional \eqref{GPE} to second order in deviations 
from the ground-state solution \eqref{bs},  as in 
\cite{lewenstein}, e.g. 
Diagonalization of the quadratic Hamiltonian 
results in the eigenvalue problem for the non-hermitian operator
\bea
{\cal L}=\left(\begin{array}{cc} 
-\frac{1}{2}\partial^2_z -2|\phi_0|^2-\mu 
& 
-\phi_0^2
\\
\phi_0^{*2}
&
\frac12\partial^2_z +2|\phi_0|^2+\mu
\end{array}\right).
\eea
Its right eigenvectors 
$(u_n,v_n)$ and corresponding adjoint modes
$(u_n^\text{ad},v_n^\text{ad})$ build a 
basis that spans the functional space of $(\phi,\phi^*)$. For all non-zero eigenvalues
$E_n$, the adjoint modes are left eigenvectors of $\cal L$. 
This is no longer true for the zero-energy modes. 
The first zero mode 
$(u_\theta,v_\theta)=i\partial_\theta(\phi_0,\phi_0^*)$, 
is related to the global $U(1)$ gauge invariance 
$\phi e^{-i\theta}\rightarrow\phi e^{-i(\theta+\epsilon)}$ 
broken by the classical solution (\ref{bs}) and its adjoint mode is 
well known,
$(u_\theta^\text{ad},v_\theta^\text{ad})=\partial_N(\phi_0,\phi_0^*)$ 
\cite{lewenstein,dziarmaga04}. 
The other zero mode, $(u_q,v_q)=i\partial_q(\phi_0,\phi_0^*)$, 
originates from the translational invariance $q\rightarrow q+\epsilon$ 
also broken by the soliton solution. To find the adjoint mode,   
one solves 
${\cal L} (u_q^\text{ad},v_q^\text{ad})
=M^{-1} 
(u_q,v_q)$, 
where $M$ is determined by the requirement 
$\la u_q^\text{ad}| u_q\ra-\la v_q^\text{ad}| v_q\ra =1$.
This ensures that $(u_q^\text{ad},v_q^\text{ad})$ is orthogonal to all 
eigenvectors of $\cal L$ with $E_n\ne 0$ \cite{lewenstein,dziarmaga04}. It is easy to verify that
\be
\left(\begin{array}{c} 
u_q^\text{ad}
\\
v_q^\text{ad}
\end{array}\right)
=
i\frac{z-q}{N}
\left(\begin{array}{c} 
\phi_0
\\
-\phi_0^*
\end{array}\right),
\ee
and $M=N$, the mass of the system.

Following Dziarmaga~\cite{dziarmaga04} we may now perform an expansion
around the ground state,
\be
\phi
=
\phi_0+
P_\theta u_\theta^\text{ad}
+P_q u_q^\text{ad}
+\sum_{n,E_n>0} \left(b_n u_n + b_n^* v_n^*\right),
\label{exp}
\ee
where all modes are implicit functions of $\theta$ and $q$.
Inserting (\ref{exp}) in the energy functional \eqref{GPE}
and expanding to second order in $P_\theta$, $P_q$ and $b_n$ 
(requiring $P_q\hl\ll N$ for the CM momentum)
results in the Bogoliubov Hamiltonian, whose quantum version reads
\be
\hat H_0=-\frac{N}{8}\hat P_\theta^2+\frac{\hat P_q^2}{2N}+\sum_{n,E_n>0}
E_n\hat b_n^\dagger \hat b_n,
\label{h0}
\ee
where $\hat P_\theta=\hat N-N=-i\partial_\theta$ and $\hat
P_q=-i\partial_q.$

Because $\theta$ can take non-perturbatively large values,
we can work in a subspace of Hilbert space with definite $\hat N=N$.
The minus sign in front of $P_\theta^2$ in \eqref{h0} arises because
the bright soliton \eqref{bs} is a saddle point of the energy functional (\ref{GPE}).
 It has no consequences since  the number of particles is fixed.

When the energy functional is supplemented with a potential term $\int dz V(z)|\phi|^2$, 
this perturbation commutes with $\hat N$, so that we can focus on its incidence on the second and third term in (\ref{h0}), 
i.e., the CM motion and Bogoliubov excitations. 
A smooth external potential can only {slightly} distort the soliton. 
{If} the potential energy variation across the soliton   
is of the order of the variance $|V_0|$ of the disordered potential, 
then 
$|V_0|\ll |\mu| = N^2/8$ is  
a sufficient condition for the soliton
shape to be only 
{weakly modified by the external potential} \cite{delta}.
Moreover, the large energy gap $|\mu|$ above the ground state \cite{ueda}  
makes Bogoliubov excitations by the potential negligible. 
{The soliton shape thus follows adiabatically and reversibly, i.e. without heating, the variations of the external
potential.}  
Then, the only degree of freedom
affected by the external potential is the soliton position $q$. 
Inserting the expansion \eqref{exp} into the energy functional,  
expanding up to quadratic terms, and 
quantizing, we arrive at  the effective Hamiltonian \eqref{bsenergy} to leading order in $1/N$ and $V$.

In the following, we study the case of 
$N=100$ Li$^7$ atoms with scattering 
length $a=-3\,$nm in a transverse harmonic trap with  
$\omega_\perp=2\pi\cdot 5\,$kHz. Then, units for energy, length, and time are 
$E_0=1.28\cdot 10^{-4}\hbar\omega_\perp$, $l_0=47.8\,\mu$m and
$t_0=0.25\,$s.  
A soliton \eqref{bs} of size $\hl=2/N=0.02$ ($\approx 1\,\mu$m) is initially prepared 
in a large axial harmonic trap with  
$\omega_z=100$ ($2\pi\cdot 64\,$Hz). 
The trapping potential is sufficiently small not to distort the
soliton, whose CM
occupies the ground state of a  harmonic oscillator 
with frequency $\omega_z$. When the trap is turned off, the soliton position starts its quantum dynamics  
with the corresponding momentum distribution 
 \be 
\label{pi0}
\pi_0(k)  = \frac{1}{\sqrt{\pi}\Delta k} \exp\left[-k^2/\Delta k ^2
\right], \quad \Delta k ^2 = N\omega_z, 
\ee
and begins to explore the disordered potential. 

In 1D random potentials, Anderson localization is  
generic:
{the amplitude of}
{every plane wave with wave vector $k$} 
{decreases asymptotically as}   
 $\exp\left\{-\gamma(k)|q|\right/2\}$. 
The inverse localization length $\gamma(k)$ 
can be calculated 
analytically in the weak-disorder limit  (see below) \cite{lifshits}.
{Equivalently, the energy spectrum is discrete and dense (in the limit of infinitely large systems) with
exponentially localized eigenstates.}

{Disorder} 
potentials are completely characterized by  
correlation functions 
$\overline{V(z_1)\dots V(z_n)}$ where the overbar denotes an ensemble average over disorder realizations.  
The average potential value shifts the origin of energy and can always be set to zero, $\overline{V(z)}=0$. 
The pair correlator can be written as $\overline{V(z')V(z'+z)}=V_0^2 C(z/\sigma_0)$, 
where $V_0$ measures the potential strength, and $\sigma_0$ the spatial correlation length. 
Higher-order correlations are required to fully describe non-Gaussian disorder such as the optical speckle potential considered in the following.   

{Optical} speckle 
yields a  light-shift potential $V(z)\propto\chi|E(z)|^2$ proportional
to the intensity of the light field $E(z)$ and to the atomic polarizability 
$\chi$, whose sign depends on the detuning of the external light frequency 
from the atomic resonance.   
At fixed detuning, the potential features either random peaks (the ``blue-detuned'' case) 
or wells (``red-detuned''). The potential distribution 
is asymmetric, and the importance of odd moments can be probed by comparing 
the blue- and red-detuned cases for fixed $|V_0|$. 
We use $|V_0|=8\cdot10^{-5}|\mu|=0.1$ in the following.   
A 1D 
speckle potential has the pair correlation function $C(y)=[\sin (y)/y]^2$, 
with a correlation length that can be as short as 
$0.26\,\mu$m \cite{Billy2008} or $\sigma_0=0.0056$ in our units. 
In $k$-space, the corresponding power spectrum  reads  
${\cal P}_V(k)=\pi\sigma_0V_0^2\; (1-|k \sigma_0|/2)\;\Theta(1-|k \sigma_0|/2)$. 

{Our} Hamiltonian \eqref{bsenergy} shows that the bright soliton sees a convoluted disorder potential whose
$k$-space components are the product $(N\pi k\hl/2)/\sinh(\pi k\hl/2)
\times V_k$ of Fourier components from 
soliton density and  
speckle potential.  
To second order (Born approximation) in the potential strength, the
inverse localization length \cite{lifshits} then reads   
\be 
\gamma(k) = \frac{N^4 \hl^2 \pi^3 \sigma_0 V_0^2 (1-|k\sigma_0|)}{[\sinh(\pi k\hl)]^2}\ \Theta(1-|k \sigma_0|) . 
\label{gammaBorn}
\ee 
For a short correlation range $\sigma_0\ll\hl$, the $k$-dependence due to the soliton convolution dominates 
and the bare speckle can be approximated by its white-noise limit  
${\cal P}_V(0)= \pi\sigma_0V_0^2$.
The soliton width $\hl$ takes over as the new effective correlation length scale. 
For $1/(\pi\hl) < k < 1/\sigma_0$ one can use the approximate expression 
\be
\gamma(k) \approx (2\pi N^2\hl)^2{\cal P}_V(0)\;e^{- 2\pi \hl k }.
\label{simplegamma}
\ee

\begin{figure}
\centering
\includegraphics*[width=0.9\linewidth]{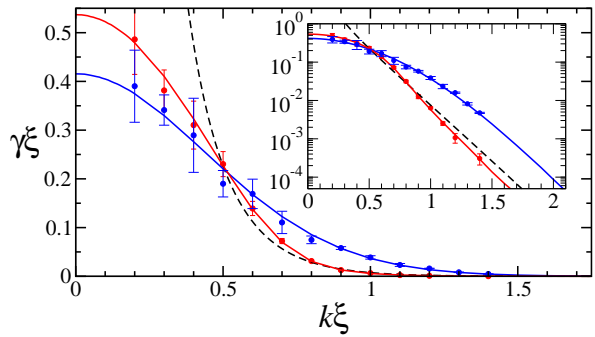}
\caption{(Color online) Inverse localization length of a bright soliton 
versus its momentum $k$ in units of its width $\hl=0.02$ 
($\approx1\,\mu$m) in a speckle potential with  
correlation length $\sigma_0=0.28\hl$. 
Circles are numerical results obtained by diagonalizing the effective Hamiltonian \eqref{bsenergy}, 
solid 
lines are numerical results from a transfer-matrix calculation.  
The bare potential amplitude is $V_0=+/-0.1$ (blue/red curve) in our units, 
corresponding to $\pm 8\cdot10^{-5}|\mu|$.   
Inset: same data in logarithmic scale. 
The lowest-order perturbation result, Eq.~\eqref{gammaBorn}, is
shown as a black dashed line. } 
\label{one}
\end{figure}

The lowest-order perturbation result \eqref{gammaBorn} 
can be compared with exact data  computed numerically,   
both by exact diagonalization and transfer matrix methods~\cite{MacKinnon1981} for the Hamiltonian \eqref{bsenergy}. 
Figure~\ref{one} {confirms that Anderson localization is observed 
for all $k$-values, but} also shows that the localization lengths for the blue- and red-detuned potential 
differ by up to an order of magnitude for $k\hl>0.5$. 
Consequently, they also differ from the lowest order perturbation results. 
Thus, perturbation theory cannot be expected to apply, 
even though the disorder potential 
is much smaller than the 
kinetic energy of the soliton. 
But the salient feature here is the rapid exponential decrease of $\gamma(k)$, 
clearly visible in the inset of
Figure~\ref{one}. Thus, 
the inverse localization lengths still obeys 
\be
\gamma=\gamma_0\exp(-\alpha k)
\label{newgam}
\ee
with $\alpha\propto\hl$, the natural length scale of the problem.

Let us return to the CM dynamics. The initial state is the superposition of 
various $k$-components, with weights given by eq.~(\ref{pi0}).
Each $k$-component localizes {at long time} into the asymptotic form
$\exp\{-\gamma(k)|q|/2\}$, implying that the superposition will also localize.
{If the extension of the initial state is small compared to the typical localization
length (which is the case here), one can neglect the phase correlations between
the various $k$-components at long time.}
The final density distribution, ensemble-averaged over disorder
realizations, is {then given by:}
\be
\label{density_int}
W(q)=\overline{|\psi(q)|^2} =  \int \rmd k \; \frac{\gamma(k)}{2} \exp\left\{-\gamma(k)|q|\right\}
\pi_0(k).
\ee 
Using \eqref{newgam} this integral can be evaluated by a saddle-point argument from which it becomes apparent that at distance $q$, 
one finds the soliton with an initial momentum $k_q$ such that 
$|q|=1/\gamma(k_q)$ provided $\alpha\Delta k\gg 1$. 
The probability distribution for the soliton position then takes the form  
\be
\label{density_approx}
W(q) = \frac{\sqrt{2}}{e \alpha\Delta k} \frac{\exp[-k_q^2/\Delta k^2]}{|q|},  
\ee  
where $k_q = \ln(\gamma_0|q|)/(\alpha)$ from \eqref{newgam}. 
In the regime of interest, this is an almost perfect algebraic decrease as $|q|^{-1}$ with a small logarithmic 
correction {\footnote{At very large distance, the exponential term in eq.~(\ref{density_approx}) becomes important, leading to
a faster decrease and eventually to a finite rms displacement $\langle q^2\rangle$ of the soliton.}}.

\begin{figure}
\centering
\includegraphics*[width=0.9\linewidth]{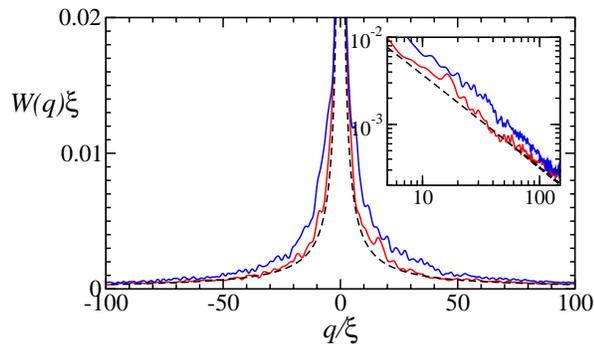}
\caption{(Color online) Probability distribution of soliton position $q$ in units of its width $\hl=0.02$ ($\approx1\,\mu$m). 
Blue and red lines: Data from numerical integration of 
the Schr\"odinger equation with the Hamiltonian (\ref{bsenergy})  starting in the ground state of an axial harmonic trap
 with $\omega_z=100$ ($2\pi\cdot 64\,$Hz) and lasting $\Delta t=20$
($5\,$s) for a blue/red-detuned speckle potential of amplitude $V_0=\pm8\cdot10^{-5}\mu$ ($2\pi\cdot 0.064\,$Hz). 
Data averaged over 250 different realizations of the disorder potential. 
Black dashed line:  saddle point result (\ref{density_approx}) for parameters in the Born approximation.
The full integral (\ref{density_int}) yields the same curve for $|q|\gg\xi$. 
The double-logarithmic inset plot shows the dominantly algebraic 
decrease as $|q|^{-1}$.}
\label{two}
\end{figure}

These predictions have been tested by numerical integration of the
Schr\"odinger equation with the Hamiltonian (\ref{bsenergy}) starting 
from the initial
Gaussian wavefunction corresponding to \eqref{pi0}, and averaging over 250 disorder realizations. 
As shown in Fig.~\ref{two}, the final probability distribution for the 
soliton position follows quite well the predicted algebraic decay. The time scale required to observe a stationary localized state around position $q$ is given by 
 $\tau(q) = |q|/v(k_q) = N |q|/k_q = \alpha|q|/\ln(\gamma_0|q|)$. 

Finally, we turn to the case of repulsive interactions~\cite{darkexp} by taking $g=+1$ in \eqref{GPE}.
{
The BEC ground state in a harmonic trap extends over the Thomas-Fermi radius. How such a condensate wave-packet expands in a disordered potential has been studied for different dimensionalities~\cite{LSP2007,Skipetrov2008,Miniatura2009}. 
Because of repulsive interaction, the entire condensate is an extended object that requires a 
field-like description, in sharp contrast with a 
bright soliton that features particle-like properties.}

There exists, however, a dark soliton solution 
$\phi_0 \propto \tanh[(x-q)/\xi]$ where $\xi=1/\sqrt{\rho}$ and $\rho$ is the background density.
Then, one can use the non-perturbative description presented in \cite{dziarmaga04} 
or a collective-coordinate method \cite{anglin} to obtain an effective Hamiltonian {for the
dynamics of the dark soliton}
in the presence of a weak disorder potential:   
\be
\label{dark}
H'_q=-\left[
\frac{P_q^2}{2|M|}+\frac{|M|}{4\xi} \int dz 
\frac{ V(z) }{\left[\cosh((z-q)/\xi)\right]^2}
\right].
\ee
The mass of the dark soliton is negative, $M=-4\rho\xi$, and its modulus 
equals twice the number of particles missing in the soliton notch.
This  effective Hamiltonian is valid only when the
velocity of the CM of the soliton is much smaller than
the sound velocity $c=\sqrt{\rho}$ in the condensate. At velocity
comparable to $c,$ the shape of the soliton changes, making the analysis more difficult.

Equation~(\ref{dark}) shows that we obtain the same form of the effective 
Hamiltonian as for the bright soliton.  
At first sight, it predicts 
similar Anderson localization effects. 
However, there is a fundamental difference between dark and bright solitons.  
The bright soliton \eqref{bs}, being the ground state of the
$N$ particle system, is protected by a large gap from quasi-particle excitations. 
In that respect, entire solitons can undergo quantum dynamics and
localize like Rubidium atoms \cite{Billy2008}.
For the dark soliton
a gap can only be imposed by the boundary conditions in 
a finite-size system, and is 
inversely proportional to the system size, 
making radiation of Bogoliubov excitations~\cite{bilas2005} and decoherence much easier. 
Furthermore, the dark soliton is an excited state, as signaled by the global 
minus sign of the effective Hamiltonian (\ref{dark}). 
Interactions with thermal cloud may  
accelerate the soliton and make it disappear \cite{shlyapnikov}.  Estimating the 
relevant time scale and evaluating its effect on the localization dynamics 
is left for future work. 

In summary we have shown that the 
center of mass of a bright soliton may undergo Anderson
localization 
in a smooth disorder potential. In realistic situations where the soliton wavepacket
is prepared in a small region of space, this leads to an {essentially} algebraic
localization of the ensemble-averaged 
atomic density at long times.
We emphasize that the effects discussed here are beyond standard
mean-field description: while the soliton's shape 
is described by a mean-field theory, 
its center of mass is treated quantum mechanically.  
This has important consequences: although the
one-body density matrix will display Anderson localization as shown in Fig.~\ref{two},
a single realization of the experiment is expected to 
find a single  
soliton at a given random position,
with a probability density given by $W(q)$, eq.~(\ref{density_approx}).

KS is grateful to Robin Kaiser for a fruitful discussion. 
Support within Polish Government scientific funds (for years 2008-2011 -- KS and 
2009-2012 -- JZ)
as a research project and by Marie Curie ToK project COCOS 
(MTKD-CT-2004-517186) is acknowledged. The research has been conducted
within LFPPI network.




\begin{thebibliography}{99}

\bibitem{McGuire64}
J.B. McGuire, J. Math. Phys. \textbf{5}, 622 (1964).

\bibitem{brightexp}
L. Khaykovich {\it et al.}, Science {\bf 296}, 1290 (2002);
K. E. Strecker {\it et al.}, Nature {\bf 417}, 150 (2002).   

\bibitem{anderson1958}
P.W.~Anderson,
Phys. Rev. {\bf 109}, 1492 (1958).

\bibitem{lee1985}
P.A.~Lee and T.V.~Ramakrishnan,
Rev. Mod. Phys. {\bf 57}, 287 (1985).

\bibitem{vantiggelen1999}
B.~van~Tiggelen, in {\it Wave Diffusion in Complex Media},
lecture notes at Les Houches 1998, edited by J.P.~Fouque, NATO
Science (Kluwer, Dordrecht, 1999).

\bibitem{Billy2008}
J.~Billy {\it et al.},
Nature {\bf 453}, 891 (2008).

\bibitem{zakharov}
V. E. Zakharov and A. B. Shabat, Sov. Phys. JETP {\bf 34}, 62 (1972).

\bibitem{lewenstein}
M. Lewenstein and L. You, Phys. Rev. Lett. {\bf 77}, 3489 (1996);
Y. Castin and R. Dum, Phys. Rev. A {\bf 57}, 3008 (1998).

\bibitem{dziarmaga04} 
J. Dziarmaga,  
Phys. Rev. A {\bf 70}, 063616 (2004).

\bibitem{Gaul2009} C. Gaul {\it et al.},
Phys. Rev. Lett. \textbf{102}, 255303 (2009).

\bibitem{delta} The upper bound on  $V_0$ excludes the case of a $\delta$-correlated 
disorder potential. 

\bibitem{ueda}
R. Kanamoto, H. Saito, and M. Ueda, Phys. Rev. A {\bf 67}, 013608 (2003).

\bibitem{lifshits}
I. M. Lifshits {\it et al.}, {\it Introduction to the Theory of Disordered Systems}
(Wiley, New York 1988). 

\bibitem{LSP2007}
L. Sanchez-Palencia \textit{et al.}, 
Phys. Rev. Lett. {\bf 98}, 210401 (2007).

\bibitem{Skipetrov2008}
S.~E.~Skipetrov, A.~Minguzzi, B.~A.~van Tiggelen, and B.~Shapiro, 
Phys. Rev. Lett. {\bf 100}, 165301 (2008).

\bibitem{Miniatura2009} 
C.~Miniatura, R.~C.~Kuhn, D.~Delande, and C.~A.~M\"uller, 
Eur.~Phys.~J.~B \textbf{68}, 353 (2009). 
 
\bibitem{MacKinnon1981}
A. McKinnon and B. Kramer, Phys. Rev. Lett. {\bf 47}, 1546 (1981).

\bibitem{darkexp}
S. Burger {\it et al.}, Phys. Rev. Lett. {\bf 83}, 5198 (1999);
J. Denschlag  {\it et al.}, Science {\bf 287}, 97, 2000.

\bibitem{anglin}
Th. Busch and J. R. Anglin, Phys. Rev. Lett. {\bf 84}, 2298 (2000);
C. K. Law, P. T. Leung and M.-C. Chu, J. Phys. B {\bf 35}, 3583 (2002).

\bibitem{bilas2005} N. Bilas and N. Pavloff,
Phys. Rev. Lett. {\bf 95}, 130403 (2005).

\bibitem{shlyapnikov}
A. Muryshev {\it et al.}, Phys. Rev. Lett. {\bf 89}, 110401 (2002);
B. Jackson, N. P. Proukakis, and C. F. Barenghi, Phys. Rev. A {\bf 75}, 051601(R) (2007). 

\end{thebibliography}
\end{document}